# Performance Evaluation of Ant-Based Routing Protocols for Wireless Sensor Networks

**Adamu Murtala Zungeru [1], Li-Minn Ang[2] and Kah Phooi Seng [3]**,

[1]School of Electrical and Electronics Engineering, University of Nottingham,
Jalan Broga, 43500 Semenyih, Selangor Darul Ehsan, Malaysia
[2]School of Engineering, Edith Cowan University,
Joondalup, WA 6027, Australia
[3]School of Computer Technology, Sunway University,
5 Jalan Universiti, Bandar Sunway, 46150 Petaling Jaya, Selangor, Malaysia

**Abstract**
High efficient routing is an important issue in the design of limited energy resource Wireless Sensor Networks (WSNs). Due to the characteristic of the environment at which the sensor node is to operate, coupled with severe resources; on-board energy, transmission power, processing capability, and storage limitations, prompt for careful resource management and new routing protocol so as to counteract the differences and challenges. To this end, we present an Improved Energy-Efficient Ant-Based Routing (IEEABR) Algorithm in wireless sensor networks. Compared to the state-of-the-art Ant-Based routing protocols; Basic Ant-Based Routing (BABR) Algorithm, Sensor-driven and Cost-aware ant routing (SC), Flooded Forward ant routing (FF), Flooded Piggybacked ant routing (FP), and Energy-Efficient Ant-Based Routing (EEABR), the proposed IEEABR approach has advantages in terms of reduced energy usage which can effectively balance the WSN node's power consumption, and high energy efficiency. The performance evaluations for the algorithms on a real application are conducted in a well known WSN MATLAB-based simulator (RMASE) using both static and dynamic scenario.

*Key words:*
Wireless Sensor Network, Energy efficiency, Performance Evaluation, Ant based routing.

## 1. Introduction

The advancement in technology has produced the availability of small and low cost sensor nodes with the integrated capability of physical sensing, data processing, and wireless communication [1-5, 20]. The decrease in the size and cost of sensors resulting from such technological advances has fueled interest in the possible use of a large set of disposable unattended sensors. But traditionally, attention has been given towards the design and development to the maximization of performance issues observed by the end users in terms of perceived throughput, quality of service (QoS), and latency. The rate of advancement in battery technology powering the sensor nodes continues to lag behind that of the semiconductor technology. The imbalance in the rate of advance which has created a gap between the energy requirements of the sensor nodes and the battery capacity that powers the nodes, calls for the design of energy-aware routing protocols so as to manage the available energy of the nodes. Sensor nodes have limited battery capacity and they must work for a satisfactory period of time. Energy is consumed by the nodes in their sensing, processing and communication tasks. Processing and communication energy consumption depends not only on the hardware, but also on the way data is routed from nodes to the sink [6].

In recent years, several competitive efficient routing algorithms for WSNs have been developed and surveyed [7-12, 21]. Recent trends in wireless sensor network routing have been towards strengthening existing approaches by considering more detailed network properties. Early work sought to adapt only the network topology such as finding a shortest path. However, WSN environment is affected by many more factors than simply changes in topology. Additional factors may include traffic congestion, latency, link quality, relative node mobility, and most importantly minimum energy path. Swarm intelligence based routing which utilizes the behavior of real biological species searching for food through pheromone deposition while dealing with problems that need to find paths to goals through the simulating behavior of ant colony finds its way in dealing with some of the challenges as mentioned above. This biologically inspired approach is proposed to adapt to the aggregate effects of each of these phenomena by finding paths of maximum throughput.

A social insect behavior suggests a probabilistic routing algorithm. Information about the network environment, including topology, link quality, traffic congestion, etc., is derived from the rate of arrival of packets at each node along with the way the respective packets generated at each node is transmitted towards the sink. This social insect environment is a representation of the network environment. Packets are considered to route themselves and are able to influence the paths of others by updating routing parameters at each node. The collection of these parameters from all nodes across the network constitutes the environment which the packets exist in. The interaction between packets and their environment implicitly spreads information about network conditions and thus reduces the need to generate explicit control traffic. The method of communicating information indirectly through the environment is known as stigmergy.

We propose a swarm intelligence based energy aware routing algorithm for wireless sensor network considering the above constraints and social insect behaviors. In this paper, we propose several improvements for EEABR [12] to increase its energy efficiency. The improvements are based on a new scheme to intelligently initialize the routing tables, giving priority to neighboring nodes that simultaneously could be the destination, intelligent update of routing tables in case of node or link failure, and reducing the flooding ability of ants for congestion control. Furthermore, the proposal maintains strong routing robustness and reliability.

The rest of the paper is organized as follows. Section 2 presents a brief review of the selected ant based routing protocols and the proposed algorithm. In section 3, we describe the simulation environment. We present our experimental and simulation results in section 4. Section 5 concludes the paper with future work intended.

## 2. A Brief Review of the Selected Ant Based Routing Protocols

### 2.1 Basic Ant Based Routing for WSN

Informally, the basic ant routing algorithm and its main characteristics [13] can be summarized as follows:

1. At regular intervals along with the data traffic, a forward ant is launched from source node to sink node.
2. Each agent (forward ant) tries to locate the destination with equal probability by using neighboring nodes with minimum cost joining its source and sink.
3. Each agent moves step-by-step towards its destination node. At each intermediate node a greedy stochastic policy is applied to choose the next node to move to. The policy makes use of (i) local agent-generated and maintained information, (ii) local problem-dependent heuristic information, and (iii) agent-private information.
4. During the movement, the agents collect information about the time length, the congestion status and the node identifiers of the followed path.
5. Once destination is reached, a backward ant is created which takes the same path as the forward ant, but in an opposite direction.
6. During this backward travel, local models of the network status and the local routing table of each visited node are modified by the agents as a function of the path they followed and of its goodness.
7. Once they have returned to their source node, the agents die.

The link probability distribution is maintained by;

$$\sum_{i \in N_k} P_{ji} = 1; \quad j = 1, \dots, N. \quad (1)$$

The traffic local model $M_k$ is updated with the values carried in $S_{s \to d}$. The trip time $T_{k \to d'}$ employed by $F_{s \to d}$ to travel from $k$ to $d'$ is used to update $\mu_d, \sigma_{d'}^2$ list $trip_k(\mu_i, \sigma_i^2)$ of estimate arithmetic mean values $\mu_i$ and associated variances $\sigma_i^2$ for trip times from node k to all nodes i (i≠k) according to the expressions:

$$\mu d' \leftarrow \mu d' + \eta (Tk \to d' - \mu d')$$

$$\sigma_{d'}^2 \leftarrow \sigma_{d'}^2 + \eta((Tk \to d' - \mu d')^2 - \sigma_{d'}^2) \quad (2)$$

The trip time $T_{k \to d'}$, $\eta$ is the weight of each trip time observed, the effective number of samples will be approximately $5(1/\eta)$,

The routing table for k is updated in the following way:
The value $P_{fd'}$ (the probability for selecting the neighbor node $f$, when the node destination is $d'$) is incremented by means of the expression:

$$Pfd' \leftarrow Pfd' + r(1 - Pfd'). \quad (3)$$

Where, $r$ is a reinforcement factor indicating the goodness of the followed path.
The $P_{nd'}$ probabilities associated to the other nodes decreases respectively:

$$P_{nd'} \leftarrow P_{nd'} - r\, P_{nd'}. \quad n \in N_k, \ n \neq f. \quad (4)$$

The factor of reinforcement $r$ is calculated considering three fundamental aspects: (i) the paths should receive an increment in their probability of selection, proportional to their goodness, (ii) the goodness is a traffic condition dependent measure that can be estimated by $M_k$, and (iii) they should not continue all the traffic fluctuations in order to avoid uncontrolled oscillations. It is very important to establish a commitment between stability and adaptability. Between several tested alternatives [14], expression (5) was chosen to calculate $r$:

$$r = c_1 \left(\frac{W_{best}}{T}\right) + c_2 \left(\frac{I_{\sup} - I_{inf}}{(I_{\sup} - I_{inf}) + (T - I_{I_{inf}})}\right) \quad (5)$$

Where $W_{best}$ represents the best trip of an ant to node $d'$, in the last observation window $W_{d'}$,
$I_{inf} = W_{best}$ stands for lower limit of the confidence interval for μ,

$I_{\sup} = \mu + z*(\sigma/\sqrt{|w|}$

Represents the upper limit of the confidence interval for µ, with

$Z = 1/\sqrt{1-\gamma}$, while $\gamma$ = confidence level, $\gamma \in [0.75, 0.8]$, $C_1$ and $C_2$ are the weight constants, chosen experimentally as $c_1 = 0.7$ and $c_2 = 0.3$ [14].

## 2.2 Sensor driven and Cost-aware ant routing (SC)

In SC [15] it is assumed that ants have sensors so that they can smell where there is food at the beginning of the routing process so as to increase in sensing the best direction that the ant will go initially. In addition to the sensing ability, each node stores the probability distribution and the estimates of the cost of destination from each of its neighbors. The protocol suffers from misleading in path discovery when there is an obstacle or lost of sight of the GPS, which might cause errors in sensing. Assuming that the cost estimate is $Q_n$ for neighbor n, the cost from the current node to the destination is 0 if it is the destination, otherwise, $C = min_{n \in N}(c_n + Q_n)$, where $c_n$ is the local cost function. The initial probability is calculated according to the expression;

$$P_n \leftarrow \frac{e^{(C-Q_n)\beta}}{\sum_{n \in N} e^{(C-Q_n)\beta}} \quad (6)$$

## 2.3 Flooded Forward ant routing (FF)

FF [15] argues the fact that ants even augmented with sensors, can be misguided due to the obstacles or moving destinations. The protocol is based on flooding of ants from source to the sink. In the case where the specific destination is not known at the beginning by the ants, or cost cannot be estimated (e.g., address-based destination), the protocol SC reduces to basic ant routing, and the problem of wandering around the network to find the destination exist. This is the case where FF exploits the network with the broadcast channel of wireless sensor networks. That is, the protocol simply uses the broadcast method of sensor networks so as to route packets to the destination. The idea is to flood forward ants to the destination. If the search is successful, forward ants will create backward ants to traverse back to the source. Multiple paths are updated by one flooding phase. Probabilities are updated in the same way as in the basic ant routing. The flooding can be stopped if the probability distribution is good enough for the data ants to the destination. The rate for releasing the flooding ants when a shorter path is traversed is reduced. Two strategies are used to control the forward flooding. First, a neighbor node will broadcast a forward ant to join the forward search only if it is closer to the destination than the node that broadcasted at an earlier time. Link probabilities are used for the estimation, i.e., a forward ant is to broadcast only if $P_n < 1/|N|$, where n is the neighbor the ant is coming from and N is the set of neighbors. If initially there is no hint, i.e., $P_n = 1/|N|$ for all n, each node will broadcast once. Secondly, delayed transmission is used in that a random delay is added to each transmission, and if a node hears the same ant from other nodes, it will stop broadcasting.

## 2.4 Flooded Piggyback ant routing (FP)

FP [15] brings a new ant species to forward ants; namely data ants whose function is to carry the forward list. The control of the flooded forward ants is the same as in FF. The protocol succeeded in combining forward ants and data ants using constrained flooding to route data and to discover optimal paths at the same time so as to minimize energy consumption of the network with the data ants carrying the forward list. In the case of control of the flooded forward ant, the data do not only pass the data to the destination, but also remember the paths which can be used by the backward ants to reinforce the probability on the links. The probability distribution constrains the flooding towards the destination for the future data ants. As compared to FF, SC, and basic ant routing in routing modeling application simulation environment (RMASE), it was found to outperforms others with high success rate, but incurred relatively high energy consumption. The method is a tradeoff between high success rate and high energy consumption.

## 2.5 Energy Efficient Ant Based Routing (EEABR)

The Energy-Efficient Ant Based Routing for WSN as proposed by T. Camilo et al. [12, 21] is an improved version of the Ant based routing in WSN. The protocol does not only consider the nodes in terms of distance but also in terms of energy level of the path traversed by the ants. The Author in his work, pointed out that, in the basic ant algorithm the forward ants are sent to no specific destination node, which means that sensor nodes must communicate with each other and the routing tables of each node must contain the identification of all the sensor nodes in the neighborhood and the correspondent levels of pheromone trail. This could be a problem since nodes would need to have a large amount of memory to save all the information about the neighborhood. In the work, the memory of the forward ant is reduced by saving only the last two visited nodes. Also proposed by the author, is the quality of a given path which should be measured based on the number of nodes on the path and the level of energy. Much improvement was observed as regards to the energy saving of the network. When compared to basic ant based routing (BABR) and improved ant based routing (IABR), it performs better in terms of energy efficiency, average energy of nodes and the energy of node with minimum energy. The disadvantages are that it lacks quality of service and increases excessive delay in packet delivery.

## 2.6 Improved Energy-Efficient Ant-Based Routing Algorithm (IEEABR)

The proposed algorithm termed Improved Energy Efficient Ant Based Routing (IEEABR) algorithm, consider the available power of nodes and the energy consumption of each path as the reliance of routing selection. It improves on memory usage, utilizes the self organization, self-adaptability and dynamic optimization capability of ant colony system to find the optimal path and multiple candidate paths from source nodes to sink nodes. The algorithm avoids using up the energy of nodes on the optimal path and prolongs the network lifetime while preserving network connectivity. This is necessary since for any WSN protocol design, the important issue is the energy efficiency of the underlying algorithm due to the fact that the network under investigation has strict power requirements. As proposed in [5], for forward ants sent directly to the sink-node, the routing tables only need to save the neighbor nodes that are in the direction of the sink-node, which considerably reduces the size of the routing tables and, in consequence, the memory needed by the nodes. As adopted in [12], the memory $M_k$ of each ant is reduced to just two records, the last two visited nodes. Since the path followed by the ants is no more in their memories, a memory must be created at each node that keeps record of each ant that was received and sent. Each memory record saves the previous node, the forward node, the ant identification and a timeout value. Whenever a forward ant is received at any node, it searches for any possible loop with the aid of its identification (ID). For the situation where no record is found, the necessary information is retrieved and the timer is restarted, hence forwarding the ant to the next node, else, the ant is eliminated if a record containing the ant identification is found. When a backward ant is received, the source ID is searched so as to know where to send it to. In this section, we proposed some modifications on EEABR to improve the Energy consumption in the nodes of WSNs and also to in turn improve the performance. The improvements are based on a new scheme to intelligently initialize the routing tables, giving priority to neighboring nodes that simultaneously could be the destination, intelligent update of routing tables in case of node or link failure, and reducing the flooding ability of ants for congestion control. The algorithm also reduces the flooding ability of ants in the network for congestion control.
The Algorithm of our proposed method is as below.

1. Initialize the routing tables with a uniform probability distribution;
$$P_{ld} = \frac{1}{N_k} \quad (7)$$
Where $P_{ld}$ is the probability $N_k$ of jumping from node l to node d (destination), the number of nodes in the network. This is done to reflect the previous knowledge about the network topology.

2. At a given time after network topology update, a greater probability values is assigned to the neighboring nodes that simultaneously could be destinations according to (8), for d ∈ $N_k$, then the initial probability in the probability distribution table of k is given by;
$$P_{dd} = \frac{9N_k - 5}{4N_k^2} \quad (8)$$
Also, for the rest neighboring nodes among the neighbors for which $m \neq d$, and $m \in N_k$ will then be:
$$P_{dm} = \begin{cases} \frac{4N_k - 5}{4N_k^2}, & if\ N_k > 1 \\ 0, & if\ N_k = 1 \end{cases} \quad (9)$$
Of course (8) and (9) satisfy (10), (note: probability distribution table is maintained by the source nodes only).

3. At regular intervals of time from every network node, a forward ant k is launched with the aim to find a path until the destination. Where the number of ants lunched at each node is limited to k*5 for network congestion control. The identifier of every visited node is saved onto a memory $M_k$ and carried by the ant. Where k is any network node having a routing table will have N entries, one for each possible destination, and d is one entry of k routing table (a possible destination). $N_k$, is the set of neighboring nodes of k, $P_{lk}$ the probability with which an ant or data packet in k, jumps to a node l, l∈$N_k$, when the destination is d d ($d \neq k$). Then, for each of the N entries in the node k routing table, it will be $n_k$ values of $P_{ld}$ subject to the condition:
$$\sum_{l \in N_k} P_{ld} = 1;\ d = 1, \dots, N \quad (10)$$

4. Forward ants selects the next hop node using the same probabilistic rule proposed in the ACO metaheuristic:
$$P_k(r,s) = \begin{cases} \frac{[\tau(r,s)]^\alpha \cdot [E(s)]^\beta}{\sum_{u \notin M_k}[\tau(r,u)]^\alpha \cdot [E(s)]^\beta}, & s \notin M_k \\ 0, else \end{cases} \quad (11)$$
where $p_k(r,s)$ is the probability with which ant k chooses to move from node r to node s, $\tau$ is the routing table at each node that stores the amount of pheromone trail on connection (r,s), E is the visibility function given by $\frac{1}{(C-e_s)}$ (c is the initial energy level of the nodes and $e_s$ is the actual energy level of node s), and α and β are parameters that control the relative importance of trail versus visibility. The selection probability is a trade-off between visibility (which says that nodes with more energy should be chosen with high probability) and actual trail intensity (that says that if on connection (r, s) there has been a lot of traffic then it is highly desirable to use that connection.

5. When a forward ant reaches the destination node, it is transformed to a backward ant which mission is now to update the pheromone trail of the path it used to reach the destination and that is stored in its memory.
6. Before backward ant k starts its return journey, the destination node computes the amount of pheromone trail that the ant will drop during its journey:

$$\Delta \tau = \frac{1}{C - \left[\frac{E_{min} - N_j}{E_{av} - N_j}\right]} \quad (12)$$

Where $C$ is the initial energy of the nodes, $E_{min}$, $E_{av}$ are the minimum and average energy respectively of the path traversed by the forward soldier as it moves towards the hill, $N_j$ represent the number of nodes that the forward soldier has visited. The idea behind the calculation of $\Delta \tau$ is that, it brings optimized routes, since it is a function of the energy level of the path, as well as length of the path. For example, a path with 10 nodes can have the same energy average as path with 4 nodes. Therefore, it is important to calculate the pheromone trail as a function of energy and number of nodes as against the number of nodes as it used in other ACO.

### 2.6.1 The Pheromone Table

The pheromone table keeps the information gathered by the forward ant. Each node maintains a table keeping the amount of pheromone on each neighbor path. The node has a distinct pheromone scent, and the table is in the form of a matrix with destination nodes listed along the side and neighbor nodes listed across the top. Rows correspond to destinations and columns to neighbors. An entry in the pheromone table is referenced by $T_{n,d}$ where $n$ is the neighbor index and $d$ denotes the destination index. The values in the pheromone table are used to calculate the selecting probabilities of each neighbor. When a packet arrives at node $B$ from previous hope $S$, i.e. the source, the source pheromone decay, and pheromone is added to link $\overrightarrow{SA}$. Route is more likely to take through $A$, since it is the shorter path to the destination i.e. $\overrightarrow{SAED}$. The pheromone table of node $A$ is shown in Figure 1 below with nodes E and S as its neighbor, B, C, E, D and S are the possible destinations. It is worth noting that all neighbors are potential destinations. At node A, the total probability of selecting links $\overrightarrow{ED}$ or $\overrightarrow{SB}$ to the destination node equal to unity (1) i.e. $\sum T_{ED} + T_{SD} = 1$. It will then be observed that, since link ED is shorter, more pheromone will be present on it and hence, route is more likely to take that path.

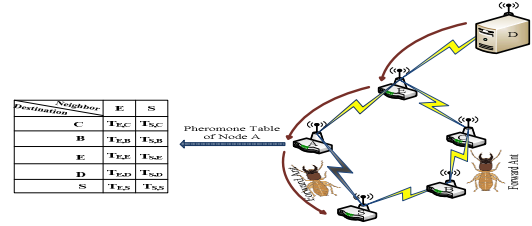

**Fig. 1** Description of pheromone table of node *A*

And the equation used to update the routing tables at each node is:

$$\tau(r,s) = (1 - \rho) * \tau(r,s) + \left[\frac{\Delta \tau}{\emptyset B d_k}\right] \quad (13)$$

Where $\phi$ a coefficient and $Bd_k$ is the distance travelled (the number of visited nodes) by the backward ant k until node r, which the two parameters will force the ant to lose part of the pheromone strength during its way to the source node. $\rho$, is a coefficient such that $(1- \rho)$ represents the evaporation of pheromone trail since the last time $\tau(r,s)$ was updated. The idea behind the behavior is to build a better pheromone distribution (nodes near the sink node will have more pheromone levels) and will force remote nodes to find better paths. Such behavior is important when the sink node is able to move, since pheromone adaptation will be much quicker.

7. When the backward ant reaches the node where it was created, its mission is completed and the ant is eliminated.
8. Else, if it fails to reach the node where it was created, i.e. when a loop is detected, immediately the ant is self destroyed.
By performing this algorithm for several iterations, each node will be able to know which are it best neighbors to send a packet towards a specific destination.
9. Whenever there is a link failure, an automatic update is made on the routing tables in case of a node n loses its link $l_{nm}$ with its neighbor node m. It is assumed that if an ant is in n, the probability $P_{dm}$, to a destination d through node m, is distributed uniformly between the remaining $N_k$-1 neighbors for the entry d in the routing table of n.

$P_{dm}$=0, during a link $l_{nm}$ failure, hence it is not possible to travel from k to m for arrival to d. Hence, new probability values after link $l_{nm}$ failure is introduce as $P_{dl}$, and the probabilities will be proportional to their relative values before the failure instead of forgetting what it has learned until the moment of the failure and is updated according to (14) as:

$$P_{dl} = P_{dl} * (1 + z) \qquad l \neq m, and\ l, m \in N_k \quad (14)$$
And,
$$z = \frac{P_{dm}}{1-P_{dm}} \quad (15)$$

With these improvements, the network converges faster, and better results were achieved.

The flow chart describing the action of movement of forward ant for our proposed Algorithm is as shown below in Figure 2. The backward ant takes the opposite direction of the flow chart, while updating the path transverse by the forward ant.

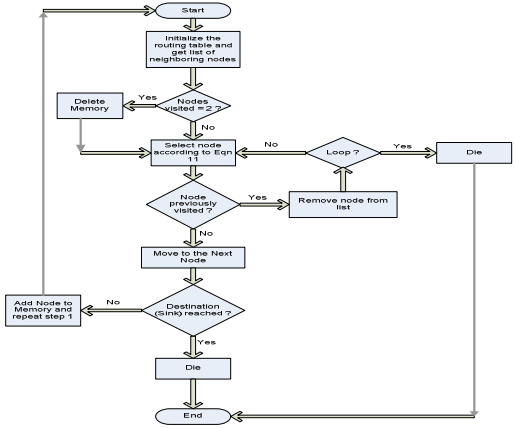

**Fig. 2** An IEEABR forward ant flow chart

## 3. Experimental and Simulation Environment

We use a Routing Modeling Application Simulation Environment (RMASE) [16] which is a framework implemented as an application in the probabilistic wireless network simulator (Prowler) [17] written and runs under Matlab, thus providing a fast and easy way to prototype applications and having nice visualization capabilities. The graphical user interface while simulating Basic ant routing is as shown in Fig. 3 below.

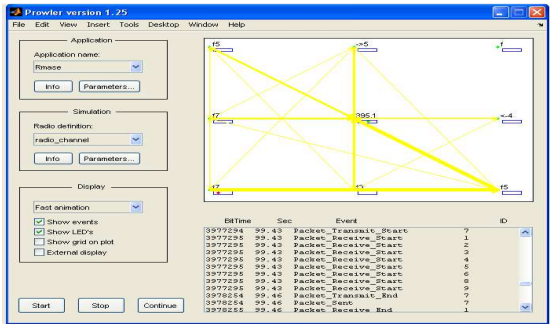

**Fig. 3** Simulation Environment showing (a) Traces of forward ants in the IEEABR routing protocol, where lines thickness indicate the probability of link selection

Prowler is an event-driven simulator that can be set to operate in either deterministic or probabilistic mode. Prowler consists of radio model as well as a MAC-layer model. The Radio propagation model determines the strength of a transmitted signal at a particular point of the space for all transmitters in the system. Based on this information, the signal reception conditions for the receivers can be evaluated and collisions can be detected. The signal strength from the transmitter to a receiver is determined by a deterministic propagation function, and by random disturbances. The transmission model is given by:

$$P_{rec,ideal}(d) = P_{transmit} \frac{1}{1+d^\gamma} \quad (16)$$

$$P_{rec}(i,j) = P_{rec,ideal}(d_{i,j}).(1 + \alpha(d_{i,j})).(1 + \beta(t)) \quad (17)$$

Where $P_{rec,ideal}$ is the ideal reception signal strength, $P_{transmit}$, the transmission signal power, d, the distance between the transmitter and the receiver, γ, a decay parameter with typical values of $2 \leq \gamma \leq 4$, α and β, random variables with normal distributions $N(0, \sigma_\alpha)$ and $N(0, \sigma_\beta)$, respectively. The MAC layer simulates the Berkeley motes' CSMA protocol, including the random waiting and back-offs.

From several results obtained from our simulation results, we report the following performance metrics for clarity purpose.

1. **Latency:** The time delay of an event sent from the source node to the destination node (seconds).
2. **Success rate:** It is a ratio of total number of events received at the destination to the total number of events generated by the nodes in the sensor network (%).
3. **Energy consumption:** It is the total energy consumed by the nodes in the network during the period of the experiment (Joules).
4. **Energy efficiency:** it is a measure of the ratio of total packet delivered at the destination to the total energy consumed by the network's sensor nodes (Kbits/Joules).

## 4. Experimental and Simulation Results

We evaluated all the protocols using the metrics defined in section 3 above. In our experiment, the network initially was a 3x3 (9) sensor grid, and later increase to 12, 36, 49, 64, and finally 100 nodes. Each experiment was performed for duration of 100 seconds. The experiment was conducted for two situations; when the sink is static, and when it is dynamic. The network of 49 nodes is generated by placing the nodes randomly in a square of 140 m x 140 m. The transmission radius of each node is set to 35 m. Other topologies are generated by scaling the square so that the average node density remains the same. The initial energy level of the nodes in the first static scenario is set to 30 J while it is 60 J in the case of the target-tracking application. The difference in energy levels is intentionally kept higher to study the energy consumption pattern of different protocols at different initial energy levels.

4.1 Static Scenario

In the static scenario, all sources and sink are fixed, while the centre of the circle is randomly selected at the start of the experiment.

**Latency:** Fig 4.a shows the end-to-end delay of the protocols under evaluation. As seen from the figure, IEEABR has the lowest end-to-end delay (latency) followed by its predecessor (EEABR). FF performance was poor, though, the basic ant routing perform worst throughout the period of observation as can be seen in the figure. The poor performance of FF and the basic ant routing is due to the flooding method of ants without control which could cause congestion in the network, hence increasing the latency. IEEABR limits the number of flooding ants in the network to a fraction of 5 times the number of networks nodes, while also assigning greater probability to neighbor who falls the same time as the sink, perform better than all the protocols.

**Success rate:** Fig 4.b shows the success rate of the protocols in other words, the ability of the protocols to deliver successfully to the sink the packets generated at each nodes in the network. Though, FP shows a wonderful performance as it delivered fully all the packets generated in the network to the sink during the period of observation without loss, where as IEEABR having an average of 96% follows. FP-Ant has the highest packet-delivery ratio followed by IEEABR in this scenario. High packet-delivery ratio of FP-Ant shows that information dissemination through flooding is robust in static networks. In this case of the converge-case scenario, the packet-delivery ratio of IEEABR is significantly higher when compared with AODV, SC, BABR, FF, and EEABR, especially in large networks. Other important observation is the poor performance of SC and the basic ant routing. The poor performance of the basic ant routing and SC is due to the flooding of ants without consideration of energy of paths, and path selection is based on distance only, in which some nodes of the paths might not be able to deliver the packets given to them for onward forwarding.

**Energy consumption:** Fig 4.c shows the energy consumption of the protocols for 9 nodes in the network. While Fig 4.e is the energy consumption of protocols for different densities of the network for the variation from 9, 16, 36, 64, and 100 nodes. SC performs better in the lower density network of 9 nodes with 3% difference in performance as against IEEABR, while IEEABR perform better when the network grows higher. Lower energy consumption of SC is due to the assumption that each node has sensors to sense the location of the sink node at the beginning of the routing process, in this case GPS. This in turns add to the cost of purchasing extra GPS to each node for practical implementation. The percentage difference between IEEABR and SC when the network grows to 49 nodes is 25%, hence, much performance difference. At that point is of 49 nodes, EEABR consumes more of 31% of energy than IEEABR. Hence outperform all the protocols in term of low energy consumption. The FP performs worst in that case as almost all the nodes went down due to high energy consumption consuming 719.9J in the network of 100 nodes where as IEEABR consumes 31.6J. The difference in the energy consumption is not comparable, even though it has the highest delivery ratio.

**Energy efficiency:** Fig 4.d shows the energy efficiency of the protocols. As energy consumption is an important metrics to be consider when designing an efficient protocol. IEEABR and EEABR are the two best protocols in terms of energy-efficiency. IEEABR better performance is due to its low total energy consumption and high packet delivery ratio. If the loss rate is high or the packet delivery rate is low as in case of BABR, it results in more route discovery processes which ultimately contribute to higher energy consumption. Another interesting observation is that FP consumes far more energy than BABR. However, their energy-efficiency figures show that BABR is close to FP which is clearly due to the poor packet-delivery rate of BABR. In this scenario, the energy-efficiency bars of IEEABR and EEABR are close to each other. On the other hand, in the target-tracking (Dynamic) application, IEEABR performs significantly better than EEABR. The reason is the ability of IEEABR to converge quickly in a dynamic scenario and achieve high packet-delivery ratio. In the static scenario, the numbers of route discoveries are very small; therefore, total energy consumption of both protocols is close to each other. However, when the number of route discoveries increases, the difference in the control-overhead gets significant contributing negatively to the energy-efficiency of EEABR. In fact virtually all the nodes ran out of energy in FF, which is the overshoot as seen in the Fig. 4(e).

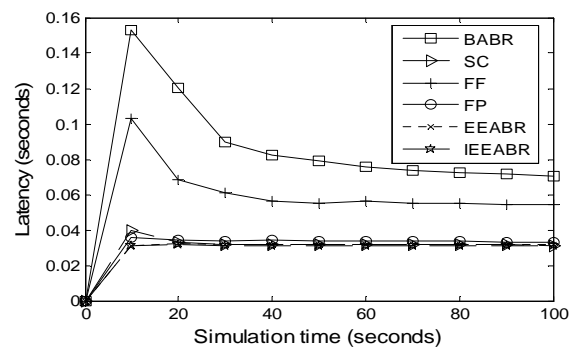

(a)

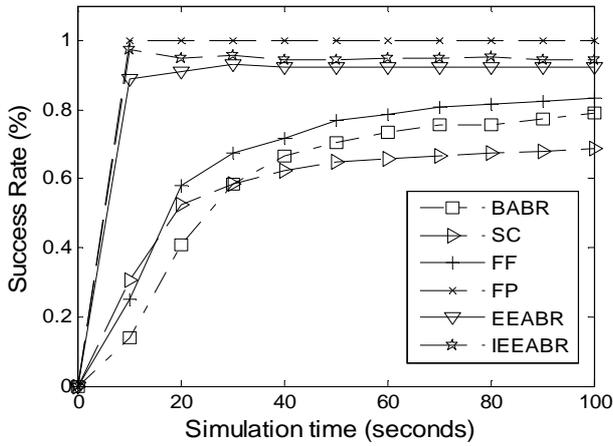

(b)

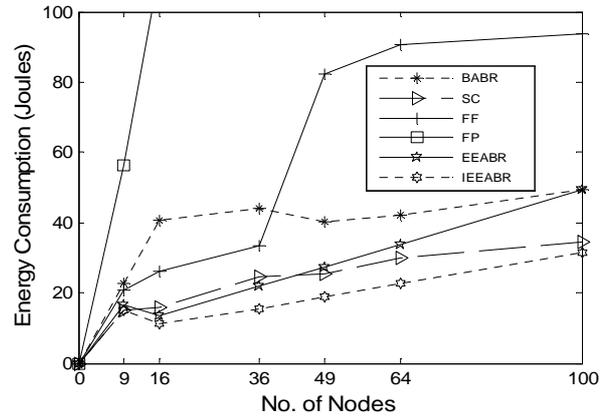

(e)

**Fig. 4** Performance evaluation in static scenario among six (6) Ant-Based routing protocols: (a) Latency (b) Success rates (c) Energy consumption (d) Energy efficiency (e) Energy consumption for different network's densities

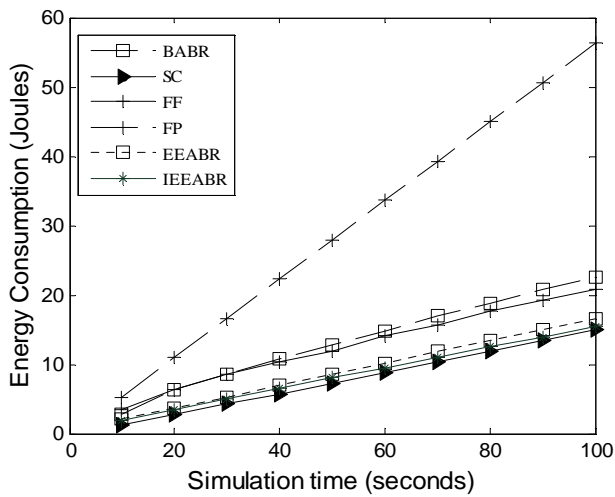

(c)

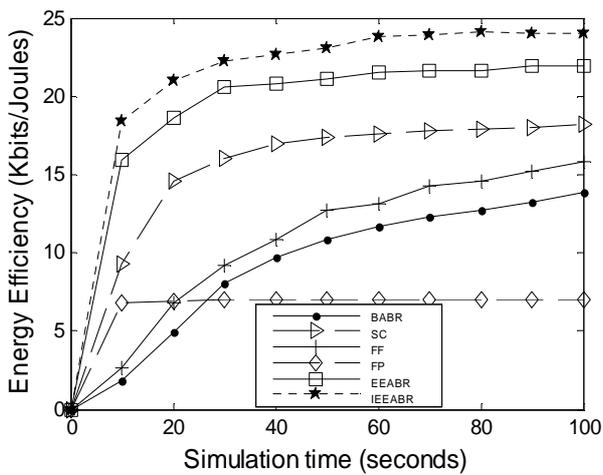

(d)

### 4.2 Dynamic Scenario

In the dynamic scenario, all source nodes are fixed while sink dynamic, and centre of the circle is randomly selected at the start of the experiment.

**Success rate:** Fig 5.a shows the success rate of the protocols in the dynamic scenario, where the sink keeps on changing position, which is sometimes known as the target tracking. The success rate of any protocol is the ability of the protocols to deliver successfully to the sink the packets generated at each node in the network. FP-Ant has the highest packet-delivery ratio followed by IEEABR in this scenario. High packet-delivery ratio of FP shows that information dissemination through flooding is more robust in dynamic networks. In this dynamic scenario, the packet-delivery ratio of IEEABR is much higher when compared with AODV, SC, BABR, FF, and EEABR, especially in large networks. Other important observation is the poor performance of SC and the basic ant routing. The poor performance of the basic ant routing and SC is due to the flooding of ants without consideration of energy of paths, and path selection is based on distance only, in which some nodes of the paths might not be able to deliver the packets given to them for onward delivery, this was also notices in the static scenario. IEEABR not only having high success rate, but also, have the lowest energy consumption and more energy efficient. it will be noticed in this scenario that IEEABR outperforms its predecessor with 60%, which is quite a large difference in performance in terms of quality of service.

**Energy consumption:** Limited available energy which is the major problem of wireless sensor networks has to be look upon critically when designing an efficient protocol.

Fig 5.b shows the energy consumption of protocols for 9 nodes in a grid network. While Fig 4.d is the energy consumption of protocols for different densities of the network for the variation from 9, 16, 36, 64, and 100 nodes. As it can be seen in Fig 5.b, SC consumes more 72.65% energy as compared to IEEABR, which shows a high performance in the static scenario, where it assumes that it knows the location of the sink using a form of sensing level or otherwise GPS to detect the position of the sink during the initial routing process. While also IEEABR shows a great improvement on EEABR with percentage difference of 10.6%. As can be seen in Fig. 5.d, the percentage difference between IEEABR and SC when the network grows to 49 nodes is 60% which is a high performance difference. IEEABR with its predecessor at that point is 29.66%. Hence outperform all the protocols in term of low energy consumption. The FP still performs worst in the tracking scenario, where almost all the nodes went down due to high energy consumption, consuming 812.7J in the network of 100 nodes where as IEEABR consumes 27.82J. The difference in the energy consumption is not comparable, even though it has the highest delivery ratio and lowest end-to-end delay in packet delivery. The high improvement is due to the reduced flooding of ants in the network, and proper initialization of the routing table, while giving preference to the sink selection among the neighbors.

**Energy efficiency:** Energy efficiency which is a function of energy consumption and the success rate, tells how well a protocol performs in both quality of service and network life time. As a network is expected to perform optimally while also performing for a long period of time without the performance degradation, Fig 5.c shows the energy efficiency of the protocols. It is clearly seen that, IEEABR not only having high success rate, low energy consumption, is the most energy efficient among the protocols under consideration. In the static converge-cast scenario, the energy-efficiency bars of IEEABR and EEABR are close to each other. On the other hand, in this target-tracking (Dynamic) application, IEEABR performs significantly better than EEABR. The reason is the ability of IEEABR to converge quickly in a dynamic scenario and achieve high packet-delivery ratio. In the static scenario, the numbers of route discoveries are very small; therefore, total energy consumption of both protocols is close to each other. However, when the number of route discoveries increases, the difference in the control-overhead gets significant contributing negatively to the energy-efficiency of EEABR. IEEABR also outperform all the routing protocols in term of Energy efficiency. The percentage difference in the dynamic scenario between IEEABR and EEABR is 64.22% and 93.2% for SC which is most costly in its algorithm implementation. FP having the highest success rate in the low density network as compared to BABR has the poorest result in term of energy efficiency.

However, their energy-efficiency figures show that BABR is close to FP-Ant which is clearly due to the poor packet-delivery ratio of BABR. Though, IEEABR and EEABR are energy aware protocols, and IEEABR still having high success rate and lowest end-to-end delay.

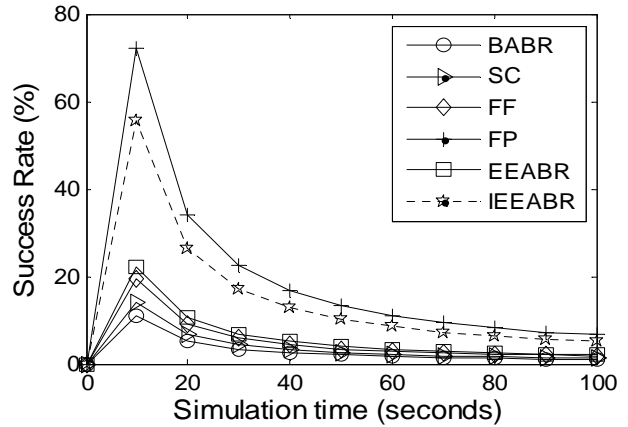

(a)

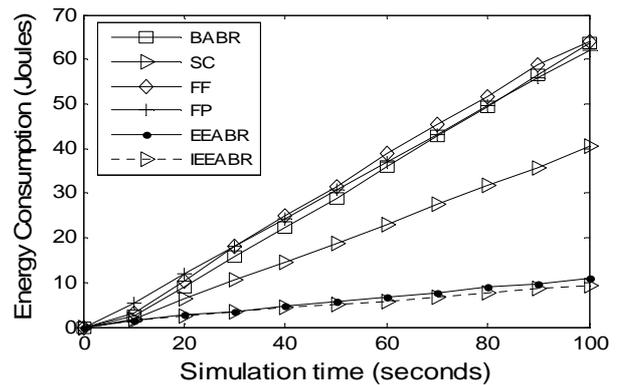

(b)

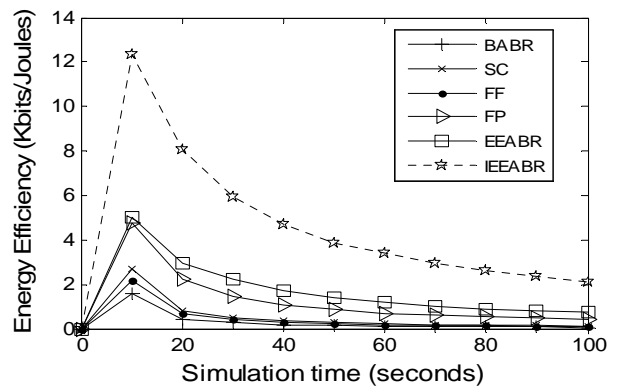

(c)

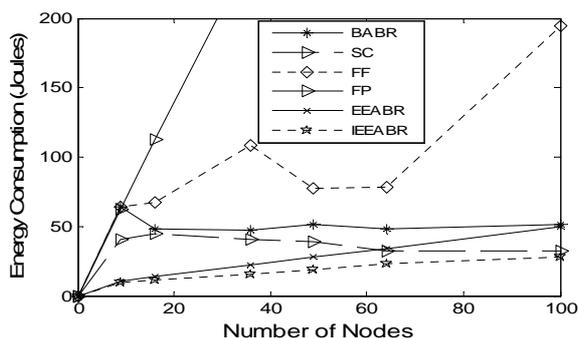

(d)

**Fig. 5** Performance evaluation of routing protocols in dynamic scenario: (a) Success rates (b) Energy consumption (c) Energy efficiency (d) Energy consumption for different network's densities

## 5. Conclusions and Future Work

In this paper, we have compared the performance of ant based routing protocols in wireless sensor networks that utilize the behavior of ants mode of communication in routing decision. we have shown that the proposed algorithm IEEABR perform quite well in all the metrics used for evaluation purpose, while also showing reasonable differences between itself and its predecessor. EEABR has 31% and 29.66% higher than IEEABR in term of energy consumption of nodes in the network for static and dynamic scenario respectively. Even SC which assumes that all sensor nodes have sensor to get the location of the sink did not do well as compared to IEEABR, despite the cost to be incurred in purchasing the GPS and attached to each sensor nodes before the performance of the algorithm. Generally, our proposed algorithm shows better performance when the network is dynamic and when the network density increases. It then shows that our proposed algorithm will do well even with low cost sensors. In future, we intend to improve on the performance due to high control packets generated as it is proactive routing algorithm. We are also on the process of implementing the algorithm on real sensor nodes (Waspmote).